\documentclass[sigconf]{acmart}
\usepackage{booktabs} 
\usepackage{bbm}
\usepackage{color, colortbl}
\usepackage{booktabs} 
\usepackage{graphicx}
\usepackage{svg}
\usepackage{courier}
\usepackage{makecell}
\usepackage{float}
\usepackage{dblfloatfix}
\usepackage{bm}
\usepackage{multirow}
\usepackage[flushleft]{threeparttable}
\usepackage{flushend}
\usepackage{bbm}
\usepackage{mathrsfs}
\usepackage{enumitem}
\usepackage{subfigure}
\usepackage{algorithm}
\usepackage{algorithmic}
\newcommand{\ignore}[1]{}
\newcommand\ChangeRT[1]{\noalign{\hrule height #1}}

\usepackage{etoolbox}
\usepackage{flushend}

\setcopyright{none}
\definecolor{xlcolor}{RGB}{218,165,32}

\newcommand{\xxx}[1]{\textcolor{red}{\bf{[CITE]}}~}

\copyrightyear{2021}
\acmYear{2021}
\setcopyright{acmcopyright}
\acmConference[KDD '21] {Proceedings of the 27th ACM SIGKDD Conference on Knowledge Discovery and Data Mining}{August 14--18, 2021}{Virtual Event, Singapore.}
\acmBooktitle{Proceedings of the 27th ACM SIGKDD Conference on Knowledge Discovery and Data Mining (KDD '21), August 14--18, 2021, Virtual Event, Singapore}
\acmPrice{15.00}
\acmISBN{978-1-4503-8332-5/21/08}
\acmDOI{10.1145/3447548.3467147}

\begin{document}
\fancyhead{}
\title{Pre-trained Language Model based Ranking in Baidu Search}
\author{
Lixin Zou,  Shengqiang Zhang$^{\dagger}$, Hengyi Cai,  Dehong Ma,
\\
Suqi Cheng,  Daiting Shi,  Zhifan Zhu, Weiyue Su, Shuaiqiang Wang,  Zhicong Cheng, Dawei Yin$^{*}$
}
\affiliation{
    \institution{Baidu Inc., Beijing, China}
}
\email{{zoulixin15,hengyi1995,chengsuqi,shqiang.wang}@gmail.com, sq.zhang@pku.edu.cn}
\email{{madehong,shidaiting01,zhuzhifan,suweiyue,chengzhicong01}@baidu.com, yindawei@acm.org}
\thanks{$^*$ Corresponding author. $^\dagger$ Co-first author.}
\begin{abstract}
As the heart of a search engine, the ranking system plays a crucial role in satisfying users' information demands. 
More recently, neural rankers fine-tuned from pre-trained language models (PLMs) establish state-of-the-art ranking effectiveness. 
However, it is nontrivial to directly apply these PLM-based rankers to the large-scale web search system due to the following challenging issues:
\textbf{(1)} the prohibitively expensive computations of massive neural PLMs, especially for long texts in the web-document, prohibit their deployments in an online ranking system that demands extremely low latency;
\textbf{(2)} the discrepancy between existing ranking-agnostic pre-training objectives and the ad-hoc retrieval scenarios that demand comprehensive relevance modeling is another main barrier for improving the online ranking system; 
\textbf{(3)} a real-world search engine typically involves a committee of ranking components, and thus the compatibility of the individually fine-tuned ranking model is critical for a cooperative ranking system.

In this work, we contribute a series of successfully applied techniques in tackling these exposed issues when deploying the state-of-the-art Chinese pre-trained language model, i.e., ERNIE, in the online search engine system. 
We first articulate a novel practice to cost-efficiently summarize the web document and contextualize the resultant summary content with the query using a cheap yet powerful Pyramid-ERNIE architecture. 
Then we endow an innovative paradigm to finely exploit the large-scale noisy and biased post-click behavioral data for relevance-oriented pre-training.
We also propose a human-anchored fine-tuning strategy tailored for the online ranking system, aiming to stabilize the ranking signals across various online components.
Extensive offline and online experimental results show that the proposed techniques significantly boost the search engine's performance.

\end{abstract}


\maketitle
{\fontsize{8pt}{8pt} \selectfont
\textbf{ACM Reference Format:}\
Lixin Zou, Shengqiang Zhang, Hengyi Cai, Dehong Ma, Suqi Cheng, Daiting Shi, Shuaiqiang Wang, Zhicong Cheng, Dawei Yin. 2021. Pre-tarined Language Model based Ranking in Baidu Search. In \text{\it Proceedings of the 27th ACM SIGKDD Conference on Knowledge Discovery and} \text{\it Data Mining (KDD '21), August 14-18, 2021, Virtual Event, Singapore.} ACM, New York, NY, USA, 9 pages. https://doi.org/10.1145/3447548.3467147}

\section{Introduction}
As essential tools for accessing information in today's world, search engines like Google and Baidu satisfy millions of users' information needs every day. In large-scale industrial search engines, \emph{ranking} typically serves as the central stage. It aims at accurately ordering the shortlisted candidate documents retrieved from previous stages, which plays a critical role in satisfying user information needs and improving user experience.

Traditional approaches, including learning to rank~\citep{Liu2010LearningTR}, are typically based on hand-crafted, manually-engineered features. However, they may easily fail to capture the search intent from the query text and infer the latent semantics of documents.
With the recent significant progress of pre-training language models (PLMs) like BERT~\citep{Devlin2019BERTPO} and ERNIE~\citep{Sun2019ERNIEER} in many language understanding tasks, large-scale pre-trained models also demonstrate increasingly promising text ranking results~\citep{Lin2020PretrainedTF}.
For example, neural rankers fine-tuned from pre-trained models establish state-of-the-art ranking effectiveness~\citep{Nogueira2019PassageRW, Nogueira2019MultiStageDR}, attributing to its ability to perform full self-attention over a given query and candidate document, in which deeply-contextualized representations of all possible input token pairs bridge the semantic gap between query and document terms.

However, it is nontrivial to directly apply the recent advancements in PLMs to web-scale search engine systems with trillions of documents and stringent efficiency requirements.
{\bf First}, significant improvements brought by these PLMs come at a high cost of prohibitively expensive computations. Common wisdom~\cite{tay2020efficient,Yang2019XLNetGA} suggests that the BERT-based ranking model is inefficient in processing long text due to its quadratically increasing memory and computation consumption, which is further exacerbated when involving the full content of a document (typically with length $>$ 4000) into the ranking stage. 
It thus poses a challenging trade-off to reconcile the efficiency and contextualization in a real-world ranking system.
{\bf Second}, explicitly capturing the comprehensive relevance between query and documents is crucial to the ranking task. Existing pre-training objectives, either sequence-based tasks (e.g., masked token prediction) or sentence pair-based tasks (e.g., permuted language modeling), learn contextual representations based on the intra/inter-sentence coherence relationship, which cannot be straightforwardly adapted to model the query-document relevance relations.
Although user behavioral information can be leveraged to mitigate this defect, elaborately designing relevance-oriented pre-training strategies to fully exploit the power of PLMs for industrial ranking remains elusive, especially in noisy clicks and exposure bias induced by the search engine.
{\bf Third}, to well deploy the fine-tuned PLM in a real ranking system with various modules, the final ranking score should be compatible with other components, such as the ranking modules of freshness, quality, authority.
Therefore, in addition to pursuing the individual performance, carefully designing the fine-tuning procedure to seamlessly interwoven the resultant PLM and other components into a cooperative ranking system is the crux of a well-behaved deployment.

This work concentrates on endowing our experiences in tackling these issues that emerged in PLM-based online ranking and introducing a series of instrumental techniques that have been successfully implemented and deployed to power the Baidu search engine. 
In order to improve both the effectiveness and efficiency axes for PLM-based full-content-aware ranking, we propose a two-step framework to achieve this goal: 
\textbf{(1)} extract the query-dependent summary on the fly with an efficient extraction algorithm; 
\textbf{(2)} decouple the text representation and interaction with a modularized PLM.
Specifically, we provide a \textbf{QU}ery-We\textbf{I}ghted Summary \textbf{E}x\textbf{T}raction (QUITE) algorithm with linear time complexity to cost-efficiently summarize the full content of the web document. 
Given a summary, a Pyramid-ERNIE, built upon the state-of-the-art Chinese PLM ERNIE~\citep{Sun2019ERNIEER},
first decouples the text representation into two parts: the query-title part and summary part. Then, the Pyramid-ERNIE captures the comprehensive query-document relevance using contextualized interactions over the previously generated representations for the sake of balancing the efficiency-effectiveness trade-off in online ranking.
To explicitly incentivize the query-document relevance modeling in pre-training Pyramid-ERNIE with large-scale raw clicking data, 
we first manage the noisy and biased user clicks through human-guided calibration by aligning the post-click behaviors with human-preferred annotations, 
and then conduct relevance-oriented pre-training using the calibrated clicks with a ranking-based objective. 
Regarding the discrepancy of ranking signals between the fine-tuned Pyramid-ERNIE and other online ranking components that emerged in the naive fine-tuning paradigm,
we alleviate such defects with a novel fine-tuning strategy in which the Pyramid-ERNIE is incentivized to be globally stabled through anchoring the fine-tuning objective with human-preferred relevance feedback, leading to better cooperation with other ranking components. 

We conduct extensive offline and online experiments in a large-scale web search engine. 
Extensively experimental results demonstrate the effectiveness of the proposed techniques and present our contributions to the relevance improvement in Baidu Search. 
We expect to provide practical experiences and new insights for building a large-scale ranking system.
Our main contributions can be summarized as follows:

\begin{itemize}[leftmargin=*]
    \item 
    \textbf{Content-aware Pyramid-ERNIE}.
    We articulate a novel practice to efficiently contextualize the web-document content with a fast query-dependent summary extraction algorithm and a Pyramid-ERNIE architecture, striking a good balance between the efficiency and effectiveness of PLM-based ranking schema in the real-world search engine system.

    \item \textbf{Relevance-oriented Pre-training}. 
    We design an innovative relevance-oriented pre-training paradigm to finely exploit the large-scale post-click behavioral data, in which the noisy and biased user clicks are calibrated to align the relevance signals annotated by the human experts.

    \item \textbf{Human-anchored Fine-tuning}. 
    We propose a human-anchored fine-tuning strategy tailored for the online ranking system, aiming to stabilize the ranking signals across various online components and further mitigate the misalignment between the naive fine-tuning objective and human-cared intrinsic relevance measurements.
    
    \item \textbf{Extensive Offline and Online Evaluations}. 
    We conduct extensive offline and online experiments to validate the effectiveness of the designed ranking approach. 
    The results show that the proposed techniques significantly boost the search engine's performance.
\end{itemize}

\section{Methodology}

In this section, we describe the technical details of our proposed approaches. 
We first formulate the ranking task as a utility optimization problem. 
Then, we provide the linear time complexity query-dependent summary extraction algorithm and propose Pyramid-ERNIE architecture to reconcile the content-aware ranking's efficiency and effectiveness.
To effectively incentivize a relevance-oriented contextual representation, we present a novel pre-training strategy in which large-scale post-click behavioral information can be distilled into the proposed Pyramid-ERNIE. 
We further design a human-anchored find-tuning schema to pertinently anchor the resulting fine-tuned model with other online ranking components.

\subsection{Problem Formulation}
The task of ranking is to measure the relative order among a set of documents $D = \left\{d_i\right\}_{i=1}^N$ under the constraint of a query $q\in \mathbb{Q}$, where $D \subset \mathbb{D}$ is the set of $q$-related documents retrieved from all indexed documents $\mathbb{D}$~\cite{liu2021plm}, and $\mathbb{Q}$ is the set of all possible queries. 
We are required to find a scoring function $f(\cdot,\cdot): \mathbb{Q}\times\mathbb{D} \rightarrow \mathbb{R}$, which can maximize some utility as
\begin{eqnarray}
f^\ast = \max_{f} \mathbb{E}_{\{q, D, Y\}} \vartheta (Y,F(q,D)). 
\end{eqnarray}
Here, $\vartheta$ is an evaluation metric, such as DCG~\citep{Jrvelin2017IREM}, PNR and ACC.
$F(q,D) = \{f(q,d_i)\}_{i=1}^N$ is the set of document scores, and $f^\ast$ is the optimal scoring function. 
$Y = \{y_i\}_{i=1}^N$ is a set of scale with $y_i$ representing the relevance label corresponds to $d_i$. 
Usually, $y_i$ is the graded relevance in 0-4 ratings, which means the relevance of $d_i$ as $\{$\textbf{bad}, \textbf{fair}, \textbf{good}, \textbf{excellent}, \textbf{perfect}$\}$ respectively.

In learning to rank, a ranking model is trained with a set of labeled query-document pairs denoted as $\Phi = \{\phi_q\}$, where $\phi_q = \{q, D = \{d_i\}, Y = \{y_i\}| 1\leq i \leq N\}$ is the set of labeled query-document given a specific query $q$. 
Under this formulation, the ranking model is learned by minimizing the empirical loss over the training data as
\begin{eqnarray}
\mathcal{L}(f) = \frac{1}{|\mathcal{Z}|} \sum_{\{q,D,Y\}\in \Phi} \ell(Y, F(q,D)), 
\end{eqnarray}
where $\ell$ is the loss function. $\ell$ is an intermediate proxy for optimizing the none-differential ranking metric $\vartheta$. $\mathcal{Z}$ is the normalizing factor. Most of the ranking models are optimized with pointwise loss (e.g., mean square error), pairwise loss (e.g., hinge loss~\citep{rosasco2004loss}), or listwise approach (e.g., LambdaMART~\citep{burges2010ranknet}).

\subsection{Content-aware Pre-trained Language Model}
In a large-scale search system, the scoring function $f(q,d)$ is typically implemented to measure the semantic relevance between the query and the document title while the document's content is ignored due to the high computational cost. 
However, merely considering the title for a query is risky since the short title usually cannot describe the document faithfully and sometimes even deviates from the content of a document, e.g., the clickbait, which presents insurmountable obstacles to the ranking effectiveness.
To incorporate the content of a document into the ranking process while allowing for fast real-time inference in a production setup simultaneously, we propose a two-step framework to achieve this goal: \textbf{(1)} we first pre-extract the query-dependent summary on the fly, which can be operated efficiently; \textbf{(2)} then, we employ a highly-modularized model---Pyramid-ERNIE, to measure the relevance of query, title, and concise summary.

\subsubsection{Query-Dependent Summary Extraction}
A document contains many contents, and correspondingly different parts may fit different queries' demand. It is more reasonable to retain the coarse-grained relevant contents and discard the rest before measuring the fine-grained semantic relevance between the query and the document.
Therefore, we propose a simple yet effective method named \textbf{QU}ery-We\textbf{I}ghted Summary \textbf{E}x\textbf{T}raction~(QUITE) to extract summary $s$ from document $d$ with respect to a given query $q$ (shown in Algorithm~\ref{alg:Framwork}). 
The QUITE first pre-processes query and documents, including word tokenization for the query, calculating the word importance, sentence segmentation for the document, and word tokenization for each sentence, respectively (line 1-4 in Algorithm~\ref{alg:Framwork}). 
Precisely, the word importance is calculated by looking up a pre-computed importance dictionary. 
Then, each sentence candidate's score $s_i$ is measured by summing the word importance of all words that appeared in both query and the sentence candidate (line 7-9 in Algorithm~\ref{alg:Framwork}). 
The candidate with the highest score will be chosen as the most related summary at the current time (line 10 in Algorithm \ref{alg:Framwork}). 
To cover different words in the summary, the importance of words that appeared in both query and the current summary will be decayed by a factor $\alpha$~$ (0 < \alpha < 1)$ (line 13 in Algorithm \ref{alg:Framwork}).
The above steps will be repeated until the number of sentences meets the predetermined threshold $k$. 
In this way, we can adaptively select the number of summaries to balance ERNIE's performance and efficiency.


\begin{algorithm}[!t] 
\caption{\small QUIET: Query-Weighted Summary Extraction} 
\label{alg:Framwork} 
\begin{algorithmic}[1]
\small
\REQUIRE ~~\\ 
The query $q$, the document $d$, the decay factor: $\alpha$;\\
The number of generated query-dependent summaries: $k$.
\ENSURE ~~\\ 
The generated query-dependent summaries: $s$;
\STATE $W_q = \text{Word-Tokenize}(q)$; 
\STATE $\omega_{w} = \text{Word-Importance}(w)$  for $w\in W_q$;
\STATE $S = \text{Sentence-Tokenize}(d)$;
\STATE $W_{s_i} = \text{Word-Tokenize}(s_i)$ for $s_i \in S$;
\STATE $s, c \gets \{\}, 1$;
\WHILE{$c \le k$}
    \FORALL{$s_i \in S$}
        \STATE $Score_{s_i} = \sum_{w\in W_o} \omega_{w}$ with $W_o = W_{s_i} \cap W_q$;
    \ENDFOR
    \STATE $s_{\ast} = {\arg\max}_{s} \{Score_{s_i}|s_i\in S \}$;
    \STATE $s \leftarrow s_{\ast} + s $; 
    \STATE $S \leftarrow S - s_{\ast} $;
    \STATE $\omega_{w} \leftarrow \alpha \cdot \omega_{w}$ for $w \in W_{s_\ast} \cap W_q$
    \STATE $c \leftarrow c + 1$;
\ENDWHILE
\RETURN $s$;
\end{algorithmic}
\end{algorithm}
\subsubsection{Pyramid-ERNIE}
\begin{figure*}
    \centering
    \includegraphics[scale=0.33]{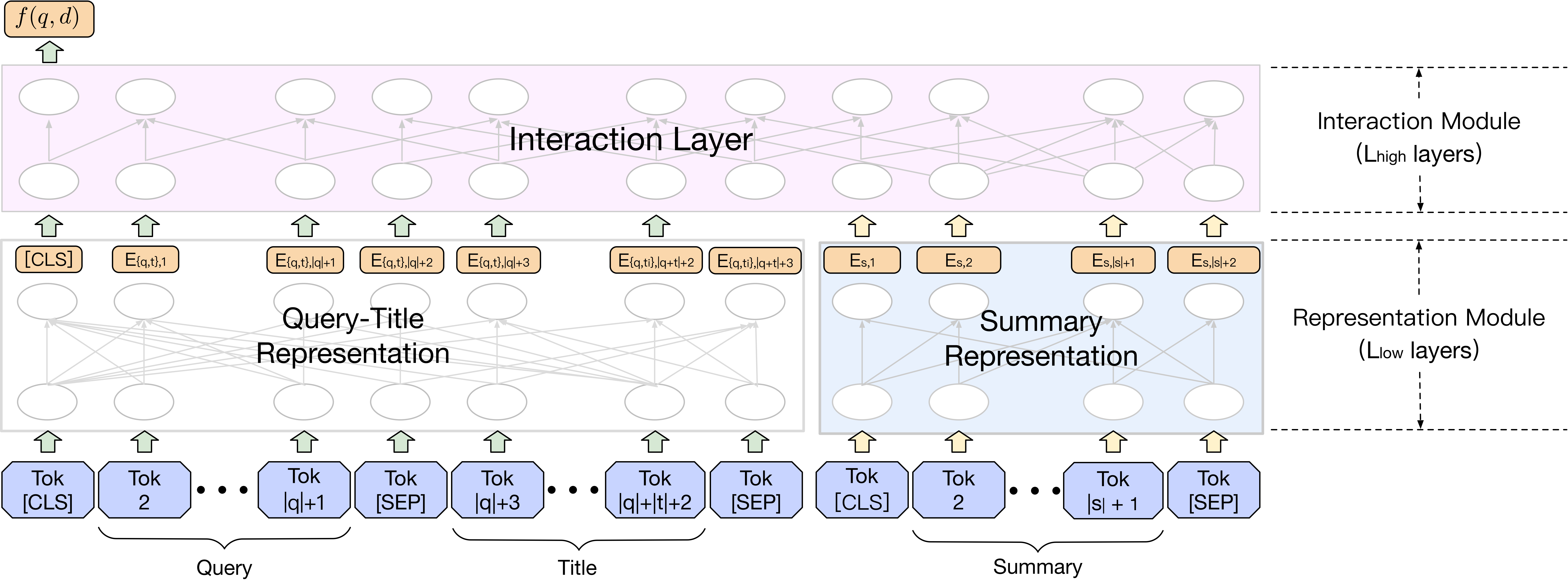}
    \caption{Illustration of the Pyramid-ERNIE model.}
    \label{fig:pyramid-ernie}
    \vspace{-5pt}
\end{figure*}

We introduce the Pyramid-ERNIE to conduct semantic matching between the query $q$, title $t$, and summary $s$.
It comprises three major components: a query and title encoder $E_{\{q,t\}} = \texttt{TRM}_{L_{low}}(q,t)$ which produces the query-title embedding, a summary encoder $E_{s} =\texttt{TRM}_{L_{low}}(s)$ which produces a summary embedding, a unified encoder $E_{\{q,t,s\}} = \texttt{TRM}_{L_{high}}(E_{\{q,t\}},E_{s})$ which encodes the concatenation of the outputs of $E_{\{q,t\}}$ and $E_s$, and produces a relevance score between the query, title and summary. 
The encoder is a $n$-layer self-attentive building block, denoted as $\texttt{TRM}_n$~(short for \textbf{TR}ansfor\textbf{M}er~\citep{Vaswani2017AttentionIA}). 
$L_{low}$ and $L_{high}$ are the number of representation layers and interaction layers respectively. 
Figure~\ref{fig:pyramid-ernie} depicts the proposed neural architecture. 

\subsubsection{Complexity Analysis}
We conduct the time complexity analysis to inspect the efficiency of the proposed approach. 
For summary extraction, the time complexity is $\mathcal{O}(N_c)$ where $N_c$ is the length of the whole content. 
Since the algorithm can be operated in linear time, the cost is rather cheap in online ranking. 
For semantic matching with Pyramid-ERNIE, the time complexity of a original ERNIE is $\mathcal{O}(Lh(N_q + N_t + N_s)^2)$, 
where $L$ and $h$ are the number of layers and hidden dimension size of ERNIE, and $N_q$, $N_t$ and $N_s$ are the length of the query, title and summaries respectively. 
In Pyramid-ERNIE, the time complexity of $E_{\{q,t\}}$, $E_{s}$ and $E_{\{q,t,s\}}$ are $\mathcal{O}(L_{low} h (N_q + N_t)^2)$, $\mathcal{O}(L_{low} h (N_s)^2)$ and $\mathcal{O}(L_{high} h (N_q+N_t+N_s)^2)$ respectively, where $L=L_{low} + L_{high}$.
Therefore, the total time complexity of Pyramid-ERNIE is $\mathcal{O}(L_{low} h (N_q + N_d)^2) + L_{low} h (N_s)^2 + L_{high} h (N_q+N_t+N_s)^2)$ which can be simplified as $\mathcal{O}(Lh(N_q + N_t + N_s)^2 - 2L_{low}h(N_q+N_t)N_s)$.
Coupled with the evidence, the time complexity of Pyramid-ERNIE is obviously lower than the original ERNIE.
This is affirmed by the empirical results, in which Pyramid-ERNIE reduces about $30\%$ time cost compared with the original ERNIE model.


\begin{figure}
    \centering
    \includegraphics[scale=0.43]{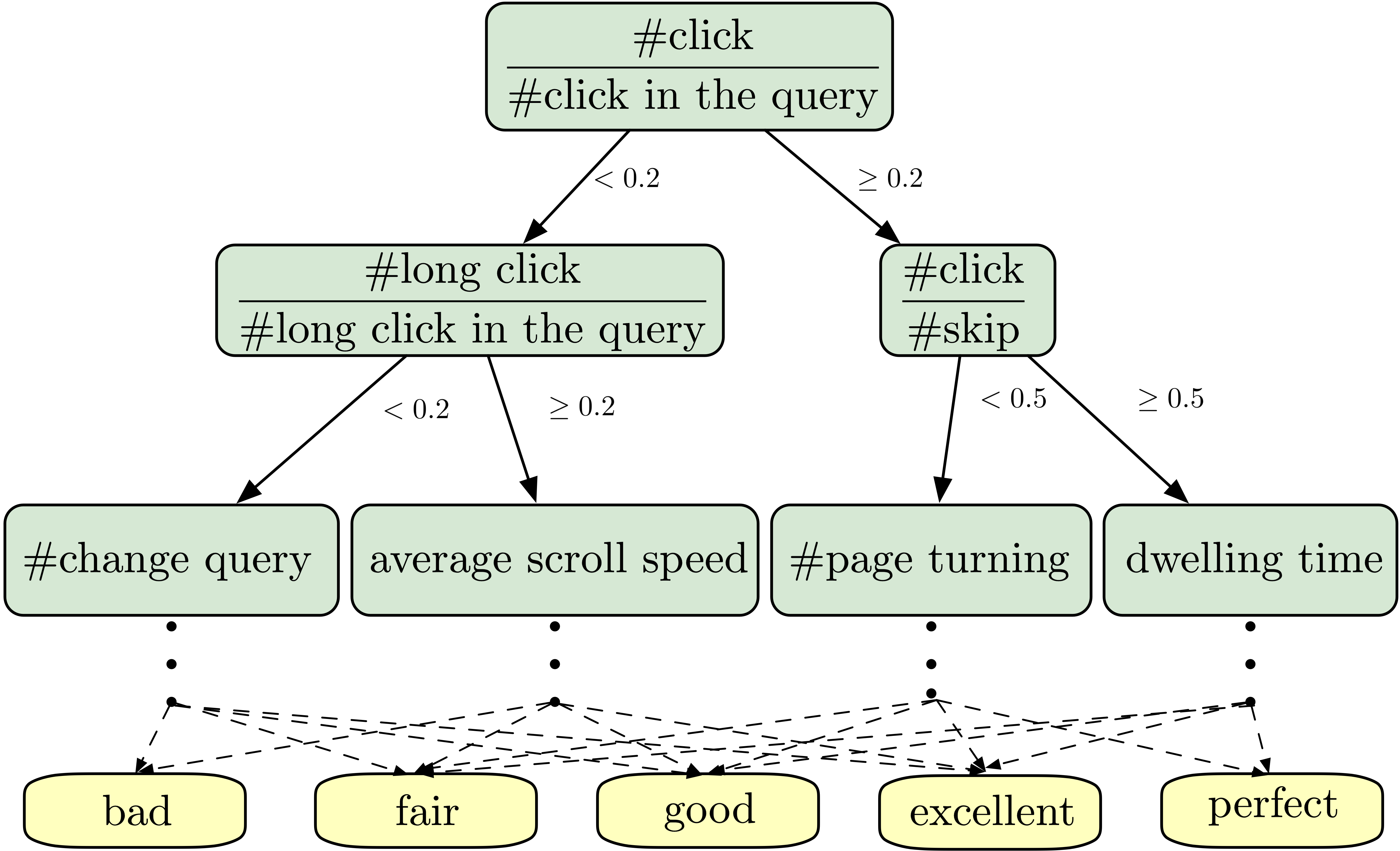}
    \caption{Human preference learning with tree-based structure.}
    \label{fig:learned_tree}
    \vspace{-5pt}
\end{figure}
\subsection{Relevance-Oriented Pre-training with Calibrated Large-Scale User Clicks}
To enhance the pre-training towards relevance ranking, one straightforward approach is to leverage the large-scale post-click behavioral information for continual domain-adaptive pre-training. The conventional training task for ranking is to predict whether the document will be clicked or not~\citep{Chapelle2009ADB}. However, such a trivial approach has the following issues:
\begin{enumerate}[leftmargin=*]
    \item 
    The clicking data contains many false-positive samples, which are caused by the noisy clicks such as clickbait and accident clicks, impairing the accurate modeling of the document relevance; 
    \item
    The exposure bias caused by the ranking system~\citep{Chen2020BiasAD} ---the displayed documents usually acquire much more clicks. It is problematic since blindly fitting the data without considering the inherent biases will result in the discrepancy between offline evaluation and online metrics, and even lead to the vicious circle of bias and rebiasing.
    \item
    The inherent inconsistency between clicking and query-documents relevance further presents obstacles to the schema of pre-training directly with the user clicking behavioral data since the documents being clicked are not necessarily relevant results.
\end{enumerate}

Fortunately, a series of informative features exhibit the fine-grained quality of user clicking, including the average dwelling time, average scroll speed, number of user-issued query rewriting and number of long-click, as well as the carefully-designed features such as $\frac{\#\text{click}}{\#\text{skip}}$, $\frac{\#\text{click}}{\#\text{click in the query}}$ and $\frac{\#\text{long click}}{\#\text{click}}$.
These important features can be leveraged to calibrate the noisy clicks and exposure bias in the raw post-click behavioral data.
For instance, the dwelling time or long-click can be used to effectively filter out the low-quality documents caused by the clickbait or accident click (issue 1);
the click skip ratio $\frac{\#\text{click}}{\#\text{skip}}$ can be employed to identify the clicks owing to exposure bias (issue 2).
To this end, we manually annotate 70 thousand query-document pairs with rich user behavioral features into 0-4 ratings and align the $M$-dimension post-click based features (denoted as $\bm{x}\in \mathbb{R}^{M}$) to query-document relevance by training a tree-based model as the calibrator to predict the human label $y$ (issue 3).
The trained tree-based calibration model can be adapted to calibrate the large-scale post-click behavioral data, and the resultant refined clicking data is finally applied to pre-train the Pyramid-ERNIE (the effectiveness of the tree-based calibration model is verified in Section~\ref{exp_REP}).
With human preference learning using a small amount of annotated data, we are able to substantially exploit the massive unsupervised data to pre-train a large ranking model and reduce the notoriously defects mentioned above.

More concretely, a classification tree~\citep{maimon2014data} $h:\mathbb{R}^{M} \rightarrow \mathbb{R}^5$ is constructed to calibrate the post-click behaviors to ground-truth relevance, as depicted in Figure~\ref{fig:learned_tree}.
Furthermore, the tree-based model is optimized using gradient boosting methods~\citep{freund1997decision}.
Note that other sophisticated classification models can also be applied, e.g., neural network-based classifier~\citep{goodfellow2016deep}.

For a given set of the query, documents, and post-click behaviors $\{q, D = \{d_i\}, X = \{\bm{x_i}\}|1\leq i \leq N\}$, we pre-train the Pyramid-ERNIE with a triplet loss defined as:
\begin{eqnarray}
\ell(G,F(q,D)) = \sum_{g(\bm{x}_i)< g(\bm{x}_j)} \max(0, f(q,d_i) - f(q,d_j) + \tau),
\end{eqnarray}
where $\tau$ is the margin enforced between positive and negative pairs, 
$g(\bm{x})=\arg\max_m \{h(\bm{x})\}_{m}$ is the most possible label generated by the tree model $h(\bm{x})$,
$G$ is the set of predicted human labels $\{g(x_i)\}_{i=1}^N$.

\subsection{Human-anchored Fine-Tuning}
Provided with a pre-trained Pyramid-ERNIE, a common practice to leverage it for online ranking tasks is to fine-tune the pre-trained Pyramid-ERNIE with the human-labeled task-specific data, using a ranking objective, e.g., pairwise loss. However, merely pursuing the individual ranking performance leads to the ranking score discrepancy between the fine-tuned Pyramid-ERNIE model and other online ranking components. This discrepancy is undesirable since a well-behaved online ranking system demands comparable ranking signals to fulfill the multi-modality and multi-source presentations of search results (e.g., freshness, authority, and quality).
Besides, optimizing the ranking model solely with pairwise loss generally suffers from the high variance of query distributions. High-relevance documents are usually overwhelmed for hot queries but extremely scarce for tail ones, posing challenges for the ranking model to perceive such cross-query relevance gap between documents.
Moreover, disregarding the query documents' intrinsic relevance also hurts the predicted scores' interpretability due to the discrepancy between the corresponding unanchored ranking scores and well-reasoned human-defined relevance grades.

Therefore, the pre-trained Pyramid-ERNIE model's final ranking score is incentivized to be globally stable across different queries and online modules by anchoring the fine-tuning objective with a human-preferred relevance judgment. 
Specifically, we manually label 10 million query-document pairs into 0-4 ratings and train the Pyramid-ERNIE with a mixture of pairwise and pointwise loss as 
\begin{equation}
  \begin{split}
  \ell(Y,F(q,D)) =& \sum_{y_i< y_j} \max(0, f(q,d_i) - f(q,d_j) + \tau) \\ 
                  & +\lambda (\delta(f(q,d_i),y_i) + \delta(f(q,d_j),y_j)),
  \end{split}
\label{eq:pointwise_plus_pairwise}
\end{equation}
where $\delta(f(q,d),y) = \max\left\{\left[f(q, d)-(\frac{y}{5}+0.1)\right]^2 - \epsilon,0\right\}$ is the pointwise loss. It endeavors to anchor the ranking score to a meaningful range, and $\epsilon = 0.01$, $\lambda = 0.7$ are the hyper-parameters.
With the pointwise loss, the ranking score $f(q,d)$ is encouraged to be consistent with the human-labeled relevance grade and can be easily blended with ranking signals from other modules in a real-world ranking system.

\section{Experiments}
To assure the effectiveness of the proposed solutions, we conducted extensive offline and online experiments on a large-scale real-world search system. 
This section details the experimental setup and presents several insights demonstrating that the proposed approaches are crucial to PLM-based ranking in a commercial search engine system.

\subsection{Dataset}
We train and evaluate our proposed method with both logged user behavioral data (\textbf{log}) and manually-labeled (\textbf{manual}) data. 
The logged data is used for the pre-training stage and the manually-labeled query-document pairs are used for the fine-tuning stage.
Specifically, we collect three months of users' accessing logs from Aug. 2020 to Oct. 2020, which contains $538,314,000$ queries and $2,986,664,000$ query-document pairs. 
Regarding the fine-tuning data, we manually annotate the train/evaluate/test dataset with Baidu's crowd-sourcing platform, resulting in $9,697,087/160,999/$ $279,128$ query-document pairs. 
In the manually-labeled training data, $73,530$ query-document pairs are used for constructing the tree-based calibrator to refine the raw user behavioral data during relevance-oriented pre-training. 
Table \ref{tab:data_statistics} offers the dataset statistics.

\begin{table}[H]
\small
\centering
\caption{Data statistics.}

\scalebox{1}{
\begin{tabular}{l|rr}
\ChangeRT{0.8pt}
Data & $\#$Query & $\#$Query-Document Pairs \\ \hline
\textbf{log} data       & 538,314,000  & 2,986,664,000 \\
\textbf{manual} train    &  469,115     & 9,697,087     \\
\textbf{manual} evaluate &  8,901       & 160,999       \\
\textbf{manual} test     &  11,437      & 279,128       \\
\ChangeRT{0.8pt}
\end{tabular}
}
\label{tab:data_statistics}
\vspace{-5pt}
\end{table}

\subsection{Evaluation Methodology}
We employ the following evaluation metrics to assess the performance of the ranking system.

The \textbf{Positive-Negative Ratio} (PNR) is a pairwise metric for evaluating the search relevance performance. 
It has been extensively used in the industry due to its simplicity. 
For a ranked list of $N$ documents, the PNR is defined as the number of concordant pairs versus the number of discordant pairs:
\begin{eqnarray}
    PNR =  \frac{\sum_{i,j\in [1,N]} \mathbbm{1}\{y_i> y_j\}\cdot \mathbbm{1}\{f(q,d_i) > f(q,d_j)\}}{
    \sum_{m,n\in [1,N]} \mathbbm{1}\{y_m > y_n\}\cdot \mathbbm{1}\{f(q,d_m) < f(q,d_n)\}},
\end{eqnarray}
where the indicator function $\mathbbm{1}\{x > y\}$ takes the value $1$ if $x>y$ and $0$ otherwise. 
We use the symbol $PNR$ to indicate this value's average over a set of test queries in our experiments.

The \textbf{Discounted Cumulative Gain} (DCG) \cite{Jrvelin2017IREM} is a standard listwise accuracy metric for evaluating the ranking model performance and is widely adopted in the context of ad-hoc retrieval. 
For a ranked list of $N$ documents, we use the following implementation of DCG
\begin{eqnarray}
    DCG_N = \sum_{i=1}^N \frac{G_i}{\log_2(i+1)},
\end{eqnarray}
where $G_i$ represents the weight assigned to the document's label at position $i$. Higher degree of relevance corresponds to the higher weight. 
We use the symbol $DCG$ to indicate the average value of this metric over the test queries. 
$DCG$ will be reported only when absolute relevance judgments are available. 
In the following sections, we will report $DCG_2$, $DCG_4$ with $N \in \{2,4\}$, respectively. 
In online experiments, we extract $6,000$ queries and manually label the top-4 ranking results generated by the search engine for calculating $DCG$.  

The \textbf{Interleaving}~\citep{chapelle2012large} is a metric used to quantify the degree of user preference and summarize the outcome of an experiment. 
When conducting comparisons with this metric, two systems' results are interleaved and exposed together to the end-users, whose clicks will be credited to the system that provides the corresponding user-clicked results. 
The gain of the new system A over the base system B can be quantified with $\Delta_{AB}$, which is defined as 
\begin{equation}
    \Delta_{AB} = \frac{wins(A) + 0.5 * ties(A, B)}{wins(A) + wins(B) + ties(A,B)} - 0.5,
\end{equation}
where $wins(\text{A})$ counts the number of times when the results produced by system A is more preferred than system B for a given query. 
Thus, $\Delta_{AB} > 0$ implies that system A is better than system B and vice versa. 
We conduct balanced interleaving experiments for comparing the proposed method with the base model.

The \textbf{Good vs. Same vs. Bad} (GSB)~\citep{Zhao2011AutomaticallyGQ} is a metric measured by the professional annotators' judgment. 
For a user-issued query, the annotators are provided with a pair (result$_1$, result$_2$) whereby one result is returned by system A, and the other is generated by a competitor system B. The annotators, who do not know which system the result is from, are then required to independently rate among Good (result$_1$ is better), Bad (result$_2$ is better), and Same (they are equally good or bad), considering the relevance between the returned document and the given query. 
In order to quantify the human evaluation, we aggregate these three indicators mentioned above as a unified metric, denoted as $\Delta \text{GSB}$:
\begin{eqnarray}
    \Delta \text{GSB} = \frac{\# \text{Good} - \# \text{Bad}}{\# \text{Good} + \# \text{Same} +  \# \text{Bad}}.
\end{eqnarray}

\subsection{Competitor System} 
Due to the high cost of deploying inferior models, we only compare the proposed method with the state-of-the-art ERNIE-based ranking model as well as different variants of the proposed approach.
\begin{itemize}[leftmargin=*]
\item Base: The base model is a $12$-layer ERNIE-based ranking policy, fine-tuned with a pairwise loss using human-labeled query-document pairs.
\item \textbf{C}ontent-\textbf{a}ware \textbf{P}yramid-ERNIE (\textbf{CAP}): This model replaces the ERNIE-based ranking model with a Pyramid-ERNIE architecture, which incorporates the query-dependent document summary into the deep contextualization to better capture the relevance between the query and document. 
\item \textbf{Re}levance-oriented \textbf{P}re-training (\textbf{REP}): This variant pre-trains the Pyramid-ERNIE model with refined large-scale user-behavioral data before fine-tuning it on the task data. 
\item \textbf{H}uman-anchored F\textbf{in}e-\textbf{t}uning (\textbf{HINT}): In the fine-tuning stage, HINT anchors the ranking model with human-preferred relevance scores using the objective function as in Equation~(\ref{eq:pointwise_plus_pairwise}). 
\end{itemize}

\subsection{Experimental Setting}
For the tree-based calibration model, we build a single tree of $6$-depth with \texttt{scikit}- \texttt{learn}~\footnote{https://scikit-learn.org/stable/}. 
Regarding Pyramid-ERNIE, we use a 12-layer transformer architecture with $9$-layers for text representation and $3$-layers for the query-title-summary interaction. It is warm-initialized with a 12-layer ERNIE 2.0 provided by Baidu  Wenxin~\footnote{https://wenxin.baidu.com/}. The $\alpha$ is set as 0.5 for query-dependent extraction.
The same hyper-parameters are used for various comparison models, i.e., vocabulary size of $32,000$, hidden size of $768$, and feed-forward layers with dimension $1024$, batch size of $128$.
We use the Adam~\citep{Kingma2015AdamAM} optimizer with a dynamic learning rate following~\citet{Vaswani2017AttentionIA}. Expressly, we set the warmup steps as 4000 and the maximum learning rate as $2\times 10^{-6}$ both in the pre-training and fine-tuning stage. 
All the models are trained on the distributed platform with $28$ Intel(R) 5117 CPU, $32G$ Memory, 8 NVIDIA V100 GPU, and 12T Disk.

\subsection{Offline Experimental Results}
Table \ref{tab_offline_result} shows the PNR results when incrementally applying the proposed techniques, i.e., CAP, REP and HINT, to the base model. The experimental result is quite consistent with our intuition. 
After adding the query-dependent summary and employing the Pyramid-ERNIE, the PNR reaches 3.017, advancing the base by 4.72\%. 
It indicates that the query-dependent summary benefits the relevance modeling, and the introduced Pyramid-ERNIE is capable of capturing the semantics of query, title, and summary. 
With the relevance-oriented pre-training, our method outperforms the base by 6.49\% and reaches the highest PNR of 3.068, which reveals that pre-training with large-scale post-click behavioral data substantially improves the performance of the ranking model. 
Finally, with the human-anchored fine-tuning strategy, although sacrificing a little bit of performance, this approach improves the stability of the Pyramid-ERNIE (referred to Section~\ref{exp_hint}). 

\begin{table}[t!]
\centering
\small
\caption{Offline comparison of the proposed methods.}
\scalebox{1}{
\begin{tabular}{l|rr}
\ChangeRT{1pt}
Model                 & PNR    & Improvement \\ \hline
Base                  & 2.881  &  -    \\ 
+CAP                 & 3.017  &  4.72\% \\
+CAP+REP           & \textbf{3.068}  &  \textbf{6.49\%} \\
+CAP+REP+HINT    & 3.065  &  6.39\% \\
\ChangeRT{1pt}
\end{tabular}
}
\label{tab_offline_result}
\vspace{-5pt}
\end{table}

\begin{table}[t!]
\centering
\small
\tabcolsep 0.01in
\caption{Performance improvements of online A/B testing.}
\scalebox{1}{
\begin{tabular}{l|cc|cc|cc}
\ChangeRT{1pt}
\multirow{2}{*}{Model} & \multicolumn{2}{c|}{$\Delta DCG$} & \multicolumn{2}{c|}{$\Delta_{AB}$ }  & \multicolumn{2}{c}{$\Delta \text{GSB}$} \\ 
& $\Delta DCG_2$ & $\Delta DCG_4$ & Random & Long-Tail & Random & Long-Tail \\\hline
Base                   &  -            &  -       &  -     &  -    &  -         &  -      \\ 
+CAP                  &  0.65\%$^\ast$ &  0.76\%$^\ast$ &  $0.15\%$  &  $0.35\%^\ast$    &  $3.50\%^\ast$       &   $6.00\%^\ast$   \\
+CAP+REP            &  2.78\%$^\ast$ &  1.37\%$^\ast$ &  \textbf{0.58}$\%^\ast$  &  $0.41\%^\ast$    &  $5.50\%^\ast$       & $7.00\%^\ast$      \\
+CAP+REP+HINT  &  \textbf{2.85\%}$^\ast$ &  \textbf{1.58\%}$^\ast$ &  0.14$\%^\ast$ &  \textbf{0.45}$\%^\ast$    &  \textbf{6.00}$\%^\ast$ & \textbf{7.50}$\%^\ast$ \\
\ChangeRT{1pt}
\multicolumn{7}{c}{$``*"$ indicates the statistically significant improvement} \\
\multicolumn{7}{c}{($t$-test with $p<0.05$ over the baseline).}\\ 
\end{tabular}
}
\label{online_ab_improve}
\vspace{-5pt}
\end{table}

\subsection{Online Experimental Results}
To investigate the effectiveness of the introduced techniques in the real-world commercial search engine, we deploy the proposed model to the online search system and compare it with the base model in the real production environment.

Table~\ref{online_ab_improve} reports the performance comparison between different models regarding $\Delta DCG$, $\Delta_{AB}$, and $\Delta GSB$. 
First, we observe that the proposed mechanisms bring substantial improvements to the online search system. 
In particular, we note that the performance of CAP, CAP+REP, CAP+REP+HINT increases gradually on $\Delta DCG_2$, $\Delta DCG_4$ and $\Delta_{AB}$ respectively, which demonstrates that the designed techniques are practical skills for improving the performance of the online ranking system. 
Moreover, we also observe that our proposed schema outperforms the online base system by a large margin for long-tail queries (i.e., the search frequency of the query is lower than 10 per week). 
Particularly, the improvements of long-tail queries in the interleaving are $0.35\%$, $0.41\%$ and $0.45\%$ for CAP, CAP+REP, CAP+REP+HINT, respectively. 
Furthermore, the advantage of GSB for the long-tail queries is $6.00\%$, $7.00\%$, and 7.50$\%$. 
We also observe that the proposed approach beats the online base system by a large margin regarding $DCG_2$ with $2.85\%$ relatively improvement. 
This reveals that the proposed schema retrieves not only relevant documents but also prefers high-quality results judged by professional annotators. 
Finally, compared with the offline experiments, we notice that the human-anchored fine-tuning strategy further boosts the online performance but slightly hurts the offline metric PNR.
This is reasonable since the human-preferred relevance annotations used in the human-anchored fine-tuning are intentionally designed to be aligned with the online users' judgments and introduced to help the ranking model cooperate with the other components, which may not well-coordinate with the offline evaluations.


\subsection{Ablation Study}
To better understand the source of the designed schema's effectiveness, we examine a series of critical strategies by analytically ablating specific parts of the proposed approach.

\subsubsection{Analysis of Content-Aware Pyramid-ERNIE}
We study different options of designing the inputs and architecture of Pyramid-ERNIE to present our motivation of setting the hyper-parameters.  

In Table~\ref{tab_pyramid_result}, we report the Pyramid-ERNIE with different settings in the interaction layers and input layers. 
As shown in Table~\ref{tab_pyramid_result}, we find that concentrating the query and title on one side and putting the summary on the other side (denoted as $qt|s$) achieves the best results. Such performance boosts can be attributed to both the early interaction between the query and title, which coarsely reflects the query and document's semantic relations and the deep interactions between the query/title and content summary, which further enrich the resulting contextualization. In contrast, coupling the title and summary on one side enables title-summary early interactions but hinders query consolidation (denoted as $q|ts$), which is crucial to the query-document relevance modeling.
As a result, the PNR consistently drops for $q|ts$ compared to $qt|s$.
Furthermore, the experiments show three layers of interaction module performs best, achieving almost equivalent performance while reducing the inference time by 25\% compared with the full self-attention-based ERNIE model. 
As expected, the performance drops when reducing the interaction module layers since insufficient interactions between query and document content make it difficult to capture the semantic relations between query and document comprehensively.

\begin{table}[t!]
    \centering
    \small
    \caption{Performance of Pyramid-ERNIE with different numbers of interaction layers. $qt|s$ denotes that the left bottom is the concatenation of $q$, $t$ and the right bottom is $s$.
    Similarly, $q|ts$ means that the left bottom is $q$ and the right bottom is the concatenation of $t$, $s$.}
    \begin{tabular}{c|cc}
        \ChangeRT{1pt}
          \# Interaction Layers &  $q|ts$ PNR  & $qt|s$ PNR \\
        \hline
             1 & 2.31  & 3.02 \\ 
             2 & 2.77  & 3.02 \\
             3 & 2.92  & 3.07 \\ 
             4 & \textbf{2.94}  & \textbf{3.07} \\ 
        \ChangeRT{1pt}
    \end{tabular}
    \label{tab_pyramid_result}
    \vspace{-5pt}
\end{table}
We explore the impact of using different summary lengths in Pyramid-ERNIE, as shown in Table~\ref{tab_qtc_result}. Without exception, increasing the number of sentences in summary leads to continuous improvement on the PNR metric.
However, a longer summary brings growing computational cost. 
Thus we select to adopt the top-1 summary as the input for Pyramid-ERNIE to balance the trade-off between efficiency and effectiveness in the large-scale online ranking system.

\begin{table}[t!]
    \centering\small
    \caption{Performance of Pyramid-ERNIE with different length of summary. $\bar{|s|}$ is the average length of summary.}
    \begin{tabular}{c|ccccc}
        \ChangeRT{1pt}
     & \makecell{w/o \\ summary}  & \makecell{1 \\ sentence} & \makecell{2 \\ sentences} & \makecell{3 \\ sentences} & \makecell{4 \\ sentences} \\\hline
PNR & 3.01 & 3.07  & 3.06 & 3.06 & \textbf{3.11}  \\
$\bar{|s|}$ & 38 & 54 & 70 & 84 & 95\\
        \ChangeRT{1pt}
    \end{tabular}
    \label{tab_qtc_result}
\end{table}
\vspace{-5pt}

\subsubsection{Influence of the Data Calibration in Relevance-Oriented Pre-training}\label{exp_REP}
As depicted in Table~\ref{tab_offline_result} and Table~\ref{online_ab_improve}, the relevance-oriented pre-training strategy effectively boosts the ranking performance.
The question then is: how do the performance improvements benefit from the tree-based calibration model?
To answer this question, we first investigate the effectiveness of the proposed tree-based calibrator.
As shown in Table~\ref{tab_pnr_upper_bound}, compared with the metric PNR estimated using raw user clicks, the proposed calibrator obtains a much higher score, indicating that the tree-based calibrator provides high-quality guidance regarding the query-document ranking relevance.
Benefiting from the refined user clicking data calibrated by this strong guidance, we further observe that pre-training with data calibration outperforms the naive pre-training strategy by a large margin, in terms of both the fine-tuning stage and the pre-training stage, as presents in Table~\ref{tab_pretraining_comparison}.
Specifically, the improvements of pre-training with calibration are 52.5\% and 23.5\% over the naive strategy.
It is worth noting that the PNR of the naive pre-training (2.83) even underperforms the base system (2.876 in Table~\ref{tab_offline_result}), affirming our intuition that the noisy and biased clicks prevalent in the user behavioral data hurt the ranking model remarkably.

\begin{table}[t!]
\centering
\small
\caption{Performance of raw user clicks and tree-based calibrator on the test set.}
        \centering
        \begin{tabular}{c|ccc}
        \ChangeRT{1pt}
        & \makecell{Raw user clicks} & \makecell{Calibrator}  \\
        \hline
        PNR & 1.86 & \textbf{3.35} \\
        \ChangeRT{1pt}
    \end{tabular}
    \label{tab_pnr_upper_bound}
    \vspace{-5pt}
\end{table}

\begin{table}[t!]
\centering
\small
\caption{Offline performance of different pre-training strategies: (a) Pre-training w/o data calibration and (b) Pre-training w/ calibrated clicking data.}
    \centering
    \begin{tabular}{c|ccc}
        \ChangeRT{1pt}
        & PNR (w/o fine-tuning) & PNR (w/ fine-tuning) \\
        \hline
        (a) & 1.81 & 2.83 \\
        (b) & \textbf{2.76} & \textbf{3.07} \\
        \ChangeRT{1pt}
    \end{tabular}
    \label{tab_pretraining_comparison}
    \vspace{-5pt}
\end{table}

\subsubsection{Effects of Human-Anchored Fine-Tuning}\label{exp_hint}
\begin{figure}[!t]
\includegraphics[width=8cm]{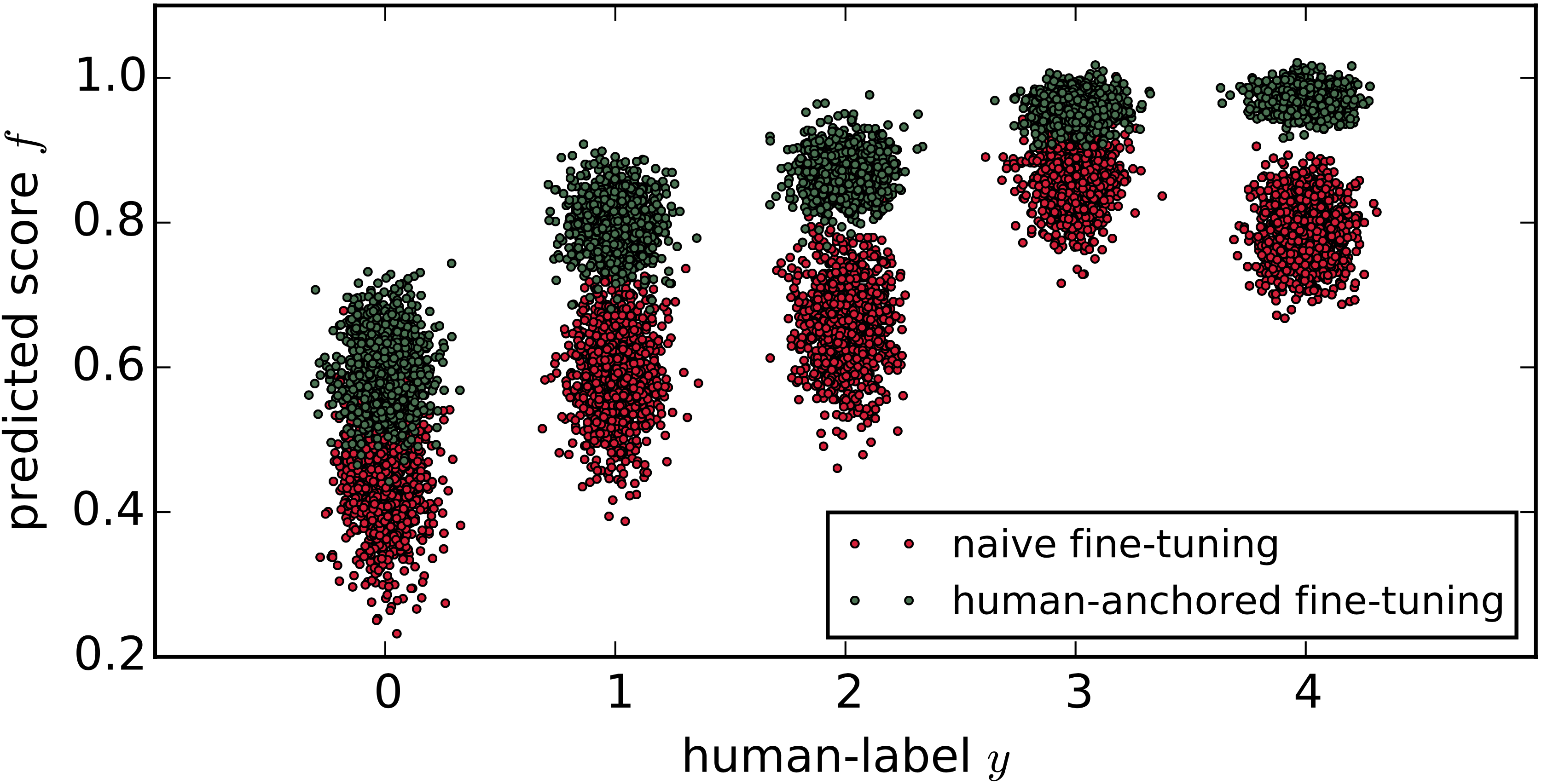}
\caption{
Scatters of prediction scores regarding the naive fine-tuning (the red dots) and human-anchored fine-tuning (the green dots) on the test set.}
\label{fig_variance}
\vspace{-5pt}
\end{figure}

In the offline and online experimental results, we show that human-anchored fine-tuning significantly improves the ranking performance at a small PNR drop cost. 
We further conduct analytical experiments to understand the source of effectiveness of this strategy.
Figure~\ref{fig_variance} scatters the relevance scores predicted by the ranking model fine-tuned with different strategies.
We notice that the human-anchored fine-tuning induces concentrated clusters around labels and lower variance of the predicted ranking scores, suggesting a more human-aligned relevance approximation in the online ranking system, which is desirable for stable and interpretable relevance estimation and online cooperation among various ranking components.
It also helps to combat the problematic cross-query relevance gap in which the query-document ranking scores are biased by the extremely long-tailed query distributions, aligning with the performance improvements of this human-anchored fine-tuning strategy in Long-Tail scenarios in online experiments (see the last line in Table~\ref{online_ab_improve}).

\section{Related Work}
\subsection{Conventional Machine Learned Ranking}
Learning-to-rank~(LTR) techniques can be categorized into three types based on the loss function: pointwise approach~\cite{cooper1992probabilistic,Li2007McRankLT}, pairwise approach~\cite{rosasco2004loss,Joachims2002OptimizingSE,Freund2003AnEB,yin2016ranking,zheng2007regression}, and listwise approach approach~\cite{burges2010ranknet,Cao2007LearningTR}. The pointwise approach, e.g., SLR~\cite{cooper1992probabilistic}, McRank~\cite{Li2007McRankLT}, assumes that each query-document pair has a relevance label, and formalizes it as a regression task. 
The pairwise approach, e.g., RankSVM~\cite{Joachims2002OptimizingSE}, RankBoost~\cite{Freund2003AnEB}, and GBRank~\cite{zheng2007regression}, treats the ranking problem as a binary classification problem and aims to learn a binary classifier discriminating which document is better in the pairwise manner. The listwise methods directly optimize ranking metrics, e.g., mean average precision, DCG/NDCG. As expected, it is better than pointwise/pairwise methods in practice. However, it is more time-consuming and difficult to be optimized. A series of listwise methods have achieved amazing results in LTR, such as LambdaRank ~\cite{burges2010ranknet}, ListNet~\cite{Cao2007LearningTR}.

\subsection{Efficient BERT-style Ranking}
Deep learning approaches have been widely adopted in the ranking, e.g., representation-based models~\cite{huang2013learning, shen2014latent}, interaction-based models~\cite{guo2016deep, xiong2017end, mcdonald2018deep,zou2020neural,zou2019reinforcement,zou2020pseudo,zhao2020whole}. 
Currently, PLM-based ranking models achieve the state-of-the-art ranking effectiveness~\cite{Nogueira2019PassageRW, Nogueira2019MultiStageDR}.
However, the performance improvement comes at the cost of efficiency since the computation cost scales quadratically to the text length. How to reconcile PLM-based ranking's efficiency and effectiveness is a seminal problem in a real-world ranking system. 
There are several research directions aiming to maintain high performance while keeping efficient computations for PLMs, including knowledge distillation~\cite{hinton2015distilling}, weight sharing~\cite{lan2019albert}, pruning~\cite{pasandi2020modeling}, and quantization~\cite{jacob2018quantization,krishnamoorthi2018quantizing}.
Besides, many works attempt to model long-text with more efficient PLMs, such as Longformer~\cite{beltagy2020longformer}, Linformer~\cite{wang2020linformer}, Reformer~\cite{kitaev2020reformer}, and Performer~\cite{choromanski2020rethinking}.
As to the ranking area, MORES~\cite{Gao2020ModularizedTR} attempts to modularize the Transformer ranker into separate modules for text representation and interaction. 
ColBERT~\cite{khattab2020colbert} introduces a late interaction architecture that independently encodes the query and the document using BERT and employs a cheap yet powerful interaction step that models their fine-grained similarity.
Our work provides a content-aware Pyramid-ERNIE architecture that balances efficiency and effectiveness in a real-world ranking system.

\subsection{Task-tailored Pre-training}
As standard PLMs usually do not explicitly model task-specific knowledge, a series of works have investigated encoding the domain knowledge into pre-trained language models.
\citet{Gururangan2020DontSP} shows that the second phase of pre-training in-domain data
leads to performance gains under both high- and low-resource settings ~\cite{Baevski2019ClozedrivenPO, Lee2020BioBERTAP, Arumae2020AnEI, Zhang2020MultiStagePF}.
To name a few, ~\citet{ma2020prop} proposes to pre-train the Transformer model to predict the pairwise preference between the two sets of words given a document;
~\citet{chang2019pre} investigates various pre-training tasks in the large-scale dense retrieval problem;
~\citet{zhang2020pegasus} designs a gap-sentences generation task as a pre-training objective tailored for abstractive text summarization;
~\citet{zhou2020pre} introduces two self-supervised strategies, i.e., concept-to-sentence generation and concept order recovering, to inject the concept-centric knowledge into pre-trained language models.
In this work, we instead perform relevance-oriented pre-training using large-scale user behavioral data and design a tree-based calibration model to refine the noisy and biased clicking data.

\subsection{Effective Fine-tuning}
Although widely adopted, existing approaches for fine-tuning pre-trained language models are confronted with issues like unstable predictions~\cite{aghajanyan2020better}, poor generalization~\cite{jiang2019smart}, or misalignment between the fine-tuning objective and designer's preferences~\cite{ziegler2019fine}.
Blindly fine-tuning the pre-trained model without considering intrinsic human-preferred task properties risks deviating the resultant fine-tuned model from human-cared ultimate goals. This paper aims to mitigate such risks by exploring a human-anchored fine-tuning strategy tailored for the online ranking system, which brings a substantial performance boost to a commercial search engine.

\section{Conclusion}
In this work, we give an overview of practical solutions to employ the state-of-the-art Chinese pre-trained language model---ERNIE---in the large-scale online ranking system. 
The proposed solutions are successfully implemented and deployed to power the Baidu search engine.
To mitigate the deficiency of existing PLMs when ranking the long web-document, we propose a novel practice to summarize the lengthy document and then capture the query-document relevance efficiently through a Pyramid-ERNIE architecture.
To manage the discrepancy between the existing pre-training objective and the urgent demands of relevance modeling in the ranking, we first provide a tree-based calibration model to align the user clicks with human preferences and then conduct the large-scale fine-tuning with refined user behavioral data.
We also articulate a human-anchored fine-tuning strategy to deal with the inconsistency of ranking signals between the Pyramid-ERNIE and other online ranking components, which further improves the online ranking performance.
The conducted extensive offline and online experiments verify the effectiveness of our proposed solutions.

\bibliographystyle{ACM-Reference-Format}
\bibliography{reference} 
\end{document}